# Early formation of moons around large trans-Neptunian objects via giant impacts

Sota Arakawa[1], Ryuki Hyodo[2], and Hidenori Genda[2]

[1]Department of Earth and Planetary Sciences, Tokyo Institute of Technology, Meguro, Tokyo, 152-8551, Japan

[2]Earth-Life Science Institute, Tokyo Institute of Technology, Meguro, Tokyo, 152-8550, Japan

**Recent studies[1,2] have revealed that all large (over 1000 km in diameter) trans-Neptunian objects (TNOs) form satellite systems. Although the largest Plutonian satellite, Charon, is thought to be an intact fragment of an impactor directly formed via a giant impact[3], whether giant impacts can explain the variations in secondary-to-primary mass ratios and spin/orbital periods among all large TNOs remains to be determined. Here we systematically perform hydrodynamic simulations to investigate satellite formation via giant impacts. We find that the simulated secondary-to-primary mass ratio varies over a wide range, which overlaps with observed mass ratios. We also reveal that the satellite systems' current distribution of spin/orbital periods and small eccentricity can be explained only when their spins and orbits tidally evolve: initially as fluid-like bodies, but finally as rigid bodies. These results suggest that all satellites of large TNOs were formed via giant impacts in the early stage of solar system formation, before the outward migration of Neptune[4], and that they were fully or partially molten during the giant impact era.**

According to the definition of the Minor Planet Center (MPC)/IAU, there are four TNOs that are classified as dwarf planets: (134240) Pluto, (136199) Eris, (136472) Makemake, and (136108) Haumea. Including these dwarf planets, there are six known TNOs with diameters larger than 1000 km (Supplementary Table 1). Thanks to the recent discovery of a moon around Makemake[1], now we know that all of these large TNOs form satellite systems.

Charon is the largest of Pluto's five satellites; the mass ratio of Charon-to-Pluto is ~ 0.12[5]. Assuming that the densities of Eris and its satellite Dysnomia are equal, their secondary-to-primary mass ratio, $\gamma_{sp}$, is ~ 0.03[6], whereas the $\gamma_{sp}$ of the Haumean system is ~ 0.0045[7]. These satellite systems are well determined for $\gamma_{sp}$. The other large TNO systems might also have similar $\gamma_{sp}$ values (on the order of $10^{-2}$ to $10^{-4}$), although information on the albedo and density of their satellites is poorly constrained.

As the largest TNOs therefore have satellites with $\gamma_{sp}$ values ranging from approximately from $10^{-1}$ to $10^{-3}$ (Supplementary Table 1), the common origin of these satellites could be related to the TNO's formation processes. Here, we focus on a giant impact onto each large TNO as a possible scenario for satellite formation. There are two mechanisms for satellite formation via a giant impact. The first is through accretion from an impact-generated debris disk, whereas the second involves direct formation from the largest intact fragment(s) of the impactor. Hereafter, we refer satellites formed via the former as "disk-origin moons" and the latter as "intact moons." Earth's moon is thought to be a disk-origin moon[8], and it is likely that Martian moons are too[9,10]. Impact simulations have shown that Charon, however is likely to be an intact moon. Several mechanisms have been proposed for the origin of the Haumean moons, also based on the giant impact scenario[11,12]. In addition, the color diversity observed among large TNOs could be explained by the variations of surface darkening as a result of giant impact-origin complex organic matter[13]. These ideas all support the theory that the satellites around these large TNOs were formed via giant impacts with variable conditions.

Here, we systematically simulated giant impacts and calculated the subsequent tidal evolutions for the satellites formed, before comparing these simulations with observational data for the large TNOs. First, we used standard smoothed particle hydrodynamics (SPH) methods to simulate the giant impacts. We then showed that giant impacts can produce satellite systems with a wide range of secondary-to-primary mass ratios, and can form both disk-origin and intact moons. We performed 434 runs for 1000 km-sized planetary bodies (see Methods), of which 141 resulted in the intact moon formation.

Figure 1 shows a typical result of a giant impact between differentiated bodies, resulting the formation of an intact moon. The impactor and target separated after the first oblique impact (Fig. 1a), but they were gravitationally bound. During the second impact, the smaller impactor received an angular momentum, and transferred its mass to the larger target (Fig. 1b). A small fragment of the impactor was then ejected because of its high specific angular momentum (Fig. 1c), leading finally to the formation of a single intact moon with $\gamma_{sp}$ = 0.13 (Fig 1d).

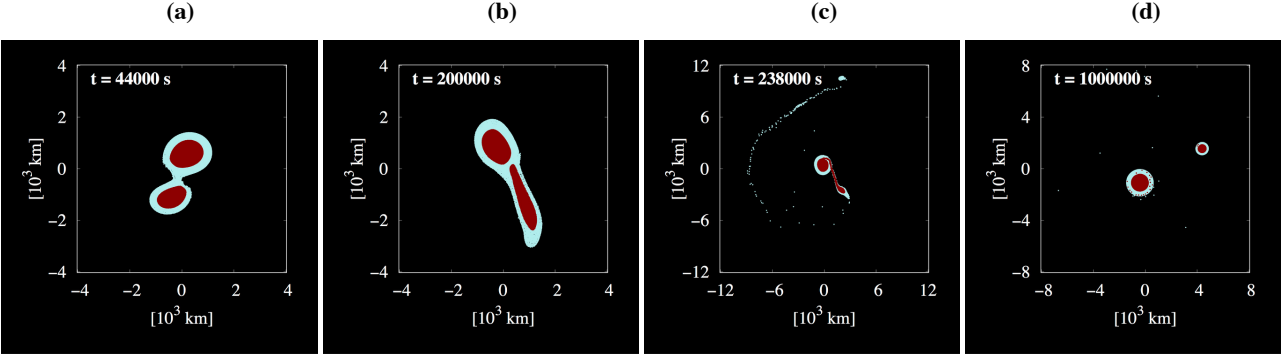

**Figure 1.** Snapshots of a giant impact between two differentiated bodies. Both the target and the impactor were differentiated bodies with ice mass fractions, $f_{ice}$, of 0.5 (50 wt.% ice and 50 wt.% basalt). The target mass, $M_{tar}$, and the impactor mass, $M_{imp}$, were $M_{tar} = 4 \times 10^{21}$ kg and $M_{imp} = 2 \times 10^{21}$ kg, respectively. The impact velocity, $v_{imp}$, was $1.05 v_{esc}$, and the impact angle, $\theta_{imp}$, was 75° (where 0° represents a head-on impact). This simulation directly formed a satellite with a $\gamma_{sp}$ of 0.13.

The collisional outcome strongly depended on $\theta_{imp}$ and $v_{imp}$, resulting in the formation of either an intact moon without a rocky core (Supplementary Fig. 1a), a disk without an intact moon (Supplementary Fig. 1b), or a hit-and-run collision (Supplementary Fig. 1c). For giant impacts between two differentiated bodies with $M_{tar} = 4 \times 10^{21}$ kg and $M_{imp} = 2 \times 10^{21}$ kg, large intact moons with $\gamma_{sp}$ values higher than $10^{-1}$ are formed when $\theta_{imp} = 75°$ and $v_{imp} = 1.05 v_{esc}$. Smaller intact moons with $10^{-3} < \gamma_{sp} < 10^{-1}$ were formed in wide parameter space in $\theta_{imp}$–$v_{imp}$ (Fig. 2a). We found that when the typical $v_{imp}$ was ≲ 1.1 $v_{esc}$, the probability of an intact moon being formed for a single giant impact reached 50%, and $\gamma_{sp}$ ranged from $10^{-3}$ to $10^{-1}$, approximately consistent with the observed $\gamma_{sp}$ ranges of the satellite systems of large TNOs.

Similarly, for the case of giant impacts with two undifferentiated icy bodies (with $M_{tar} = 4 \times 10^{21}$ kg and $M_{imp} = 2 \times 10^{21}$ kg), large intact moons with $\gamma_{sp} \simeq 10^{-1}$ formed during some grazing impacts with $\theta_{imp}$ values of 60° or larger. In addition, we also found that intact moons with $10^{-3} < \gamma_{sp} < 10^{-1}$ formed via giant impacts in wide parameter space where $\theta_{imp}$–$v_{imp}$ (Fig. 2b). These SPH simulations agree with those of previous studies[3,13], although these previous studies focused only on the formation of Charon, and therefore set the total angular momentum of the system as the current Pluto-Charon value, whereas we surveyed various $\theta_{imp}$ and $v_{imp}$ values.

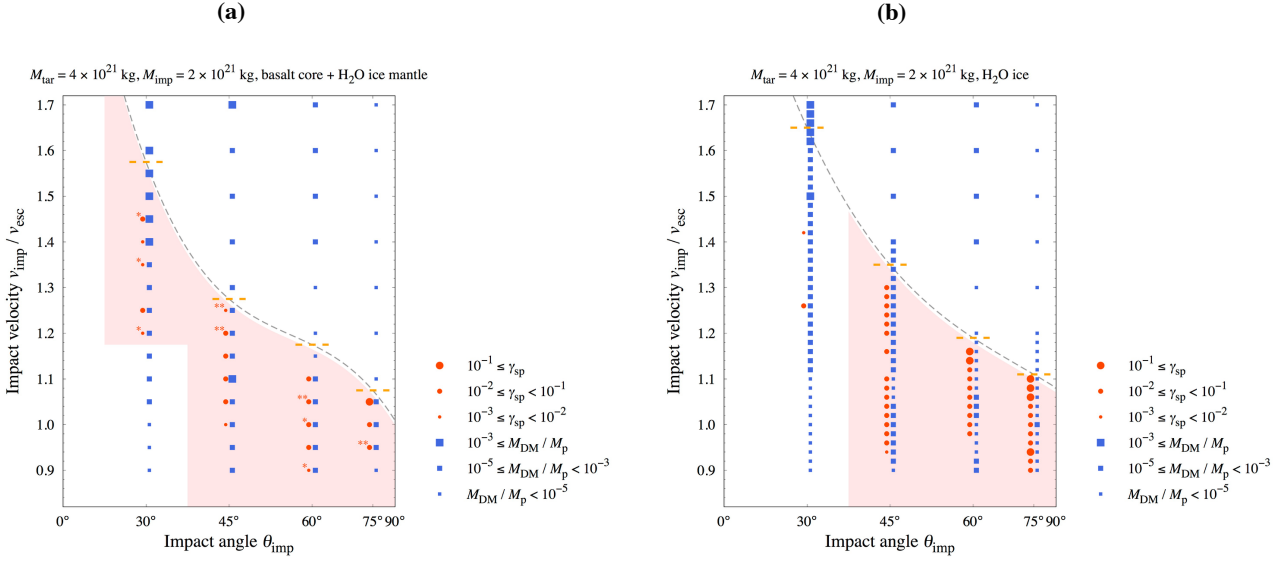

**Figure 2.** Summary of the range of outcomes for the simulated giant impacts. (a) outcomes for giant impacts between differentiated planetary bodies with masses of $4 \times 10^{21}$ kg and $2 \times 10^{21}$ kg. Both the impactor and target were differentiated with a $f_{ice}$ of 0.5. (b) outcomes for giant impacts between icy undifferentiated planetary bodies with masses of $4 \times 10^{21}$ kg and $2 \times 10^{21}$ kg. The size (large, medium, or small) of each red circle represents the mass of the intact moon (if formed), and the size of each blue square represents the estimated mass of the disk-origin moon. Asterisks signify that multiple intact moons were formed during the single impact (a single asterisk implies the formation of one large intact moon and another small intact moon, two asterisks imply the formation of one large intact moon and two other small intact moons). The orange dashed lines represent the criteria for hit-and-run collisions[14], and the grey curve is a fit calculated using a cubic function. The pink-tinted region is a rough estimation of the parameter range for intact moon formation.

We found that intact moons were formed via giant impacts when $\theta_{imp} \gtrsim 45°$ without strong dependence on whether the impactor and the target were differentiated, although that the formation probability of large intact moons where $\gamma_{sp} > 10^{-1}$ was higher than that of giant impacts with differentiated bodies. Supplementary Figs. 2a–2c show the outcomes of giant impacts for different total masses, different compositions, and different impactor-to-target mass ratios. In these conditions, intact moons were formed in almost all cases where $\theta_{imp} \gtrsim 45°$.

Assuming that the probability distribution of the impact angle, $P(\theta_{imp})$, is given by $P(\theta_{imp}) = \sin(2\theta_{imp})$, then stochastically, half of the giant impacts had an impact angle larger than 45°. Typical impact velocities among large (> 1000 km in diameter) TNOs are given by $v_{imp} \simeq ((e_{hel}^2 + i_{hel}^2) v_K^2 + v_{esc}^2)^{1/2}$, where $e_{hel}$ and $i_{hel}$ are the heliocentric eccentricity and the inclination respectively, and $v_K$ and $v_{esc}$ are the heliocentric Kepler velocity and the escape velocity of the TNOs, respectively[3]. When $e_{hel}$ and $i_{hel}$ ≪ 0.1 in the giant impact era, the relative velocity of large TNOs at infinity, $v_\infty \sim (e_{hel}^2 + i_{hel}^2)^{1/2} v_K$, will be smaller than $v_{esc}$ (~ 1 km s$^{-1}$ for large TNOs), and the impact velocity will be $v_{imp} \sim v_{esc}$. Although the current heliocentric eccentricities of large TNOs are much higher than 0.1 (e.g., currently $e_{hel}$ = 0.25 for Pluto), we note that the $v_\infty$ of large TNOs might be small during the giant impact era. This is because $e_{hel}$ and $i_{hel}$ are smaller than 0.1 for the cold classical TNOs, which were thought to be formed locally[15], and large TNOs can also be formed locally, in a similar way to the smaller cold classical TNOs[16].

In order to occur with reasonable frequency, giant impacts between large TNOs must be gravitationally focused. In our simulations, when $v_\infty$ was smaller than the Hill velocity, $v_H \sim (M_{TNO}/M_\odot)^{1/3} v_K$, the collision timescale was ~ 6 Myr, ($M_{TNO}$ and $M_\odot$ are the masses of the large TNO and the solar mass, respectively)[17]. If the $e_{hel}$ and the $i_{hel}$ of the TNOs were on the order of $10^{-3}$, then they were gravitationally focused (i.e., $v_\infty < v_H$)[16]. In this case, the typical impact velocity was ~ $1.0 v_{esc}$, consistent with the formation of intact moons via giant impacts. The small $v_\infty$ value is also suggested in the context of the formation of trans-Neptunian binaries[17].

Concerning the possibility of the formation of viscous disk-origin moons by giant impacts[18], we analyzed the disk mass around the primary at $10^6$ seconds after the start of calculation and then estimated the mass of the disk-origin moon using the empirical relation between disk mass and satellite mass (see Methods). As a result of this analysis, we found that large satellites for which $M_{DM}/M_p > 10^{-3}$ were very rarely formed via the viscous spreading of a debris disk, except for the case of high-speed impacts of $v_{imp} \gtrsim 1.4 v_{esc}$ (Figs. 2a, 2b, and Supplementary Figs. 2a−2c). This result might indicate that the secondaries around large TNOs, which have $\gamma_{sp}$ ranges from $10^{-3}$–$10^{-1}$, are intact moons, not disk-origin moons.

In our simulations, giant impacts between differentiated bodies sometimes formed multiple intact moons via a single impact (Fig. 2a). This indicates the possibility of collision between multiple intact moons. In contrast, no multiple intact moons were formed during simulations of single giant impacts between undifferentiated bodies (Figs. 2b and Supplementary Figs. 2a−2c). The question of whether multiple satellites will survive, or instead merge into single large satellite, is beyond the scope of this study. The possibility of collisions between multiple intact moons does have the potential to explain the origin of Hi'iaka, the secondary of Haumea; its spin is non-synchronous, which if Hi'iaka has not experienced further impacts since its formation, is unexpected (see Supplementary Section 3).

The $f_{ice}$ values of satellites formed after giant impacts from differentiated bodies showed a strong correlation with $\gamma_{sp}$ (Supplementary Fig. 3). The $f_{ice}$ values of the resulting intact moons were similar to that of the pre-impact targets and impactors ($f_{ice}$ = 0.5) when $\gamma_{sp} \gtrsim 10^{-2}$. In contrast, for small intact moons and disk-origin moons with $\gamma_{sp} \lesssim 10^{-2}$, $f_{ice} \simeq 1$. These values can be compared to observations on the density of both primary and secondary satellites for two systems (Pluto-Charon and Haumea-Hi'iaka). For Pluto-Charon ($\gamma_{sp}$ = 0.12), the bulk density of Pluto (1860 kg m$^{-3}$; ref. [19]) is similar to that of Charon (1700 kg m$^{-3}$; ref. [19]), suggesting that both Pluto and Charon have a similar $f_{ice}$. On the other hand, for Haumea-Hi'iaka system ($\gamma_{sp}$ = 4.5 × $10^{-3}$), the bulk density of Haumea (1800 kg m$^{-3}$; ref. [20]) is larger than that of Hi'iaka (≈ 1000 kg m$^{-3}$; ref. [8]). Therefore, the secondary, Hi'iaka, is a pure icy body, while its primary, Haumea, consists of both ice and rock. These observational constraints are consistent with the theory that both large and small satellites around large TNOs are impact moons formed via giant impacts between differentiated bodies, although we do not reject the possibility of the formation of intact moons from undifferentiated bodies.

During simulation of the formation of the intact moon, the periapsis distance $q_{ini}$ was typically 3–4$R_p$, where $R_p$ is the planetary radius of the primary (Fig. 3). The eccentricity ($e_{ini}$) was distributed across all ranges, from zero to one (Fig. 3). This is consistent with a previous study that simulated the impact that formed the Pluto-Charon system[3].

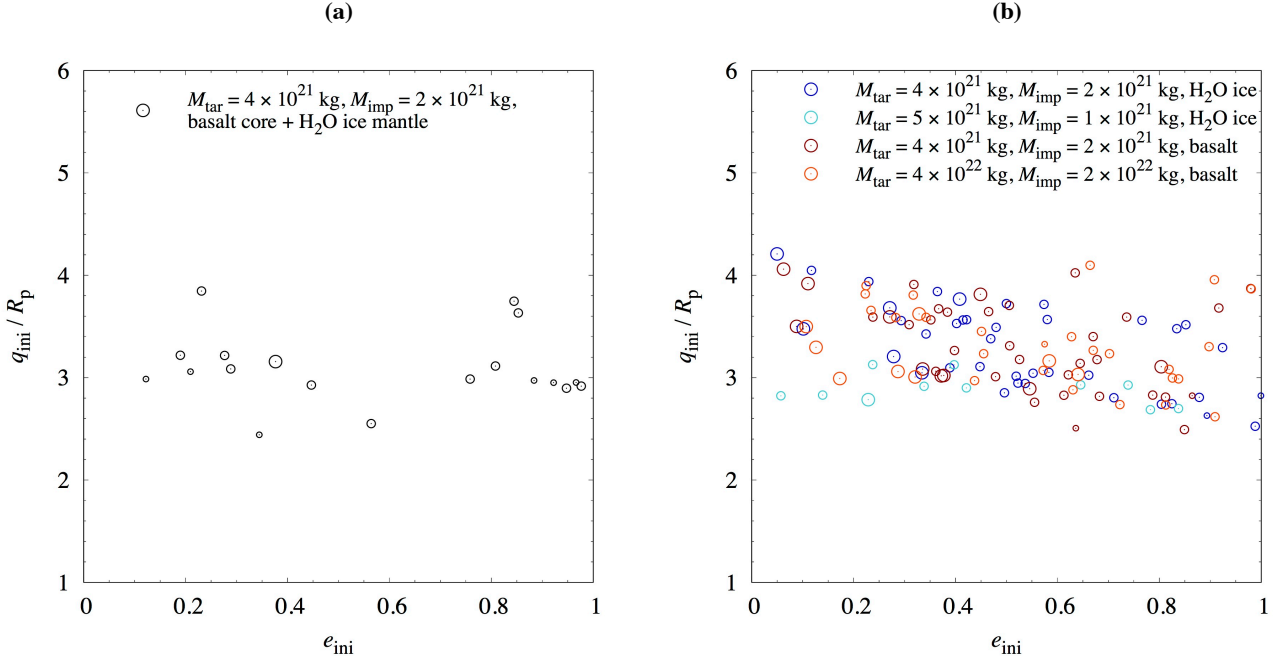

**Figure 3.** Initial distribution of periapsis distance $q_{ini}$ and eccentricity $e_{ini}$ before tidal evolution. The size (large, medium, small) of each circle represents $\gamma_{sp}$ ($10^{-1} \leqq \gamma_{sp}$ for large circles, $10^{-2} \leqq \gamma_{sp} < 10^{-1}$ for medium-sized circles, and $10^{-3} \leqq \gamma_{sp} < 10^{-2}$ for small circles). (a) the case for differentiated bodies. (b) the case for undifferentiated bodies.

We then performed semi-analytical tidal evolution calculations (see Methods) to discern whether intact moons formed via giant impacts can evolve into satellites with circular orbits. Tidal evolution was found to be strongly dependent on the material states, and we found that most of the intact moons turned into eccentric satellites when their planetary bodies always behaved as rigid bodies, while almost intact moons adopted circular orbits when their planetary bodies initially had small rigidity and behaved as fluid-like bodies (Fig. 4).

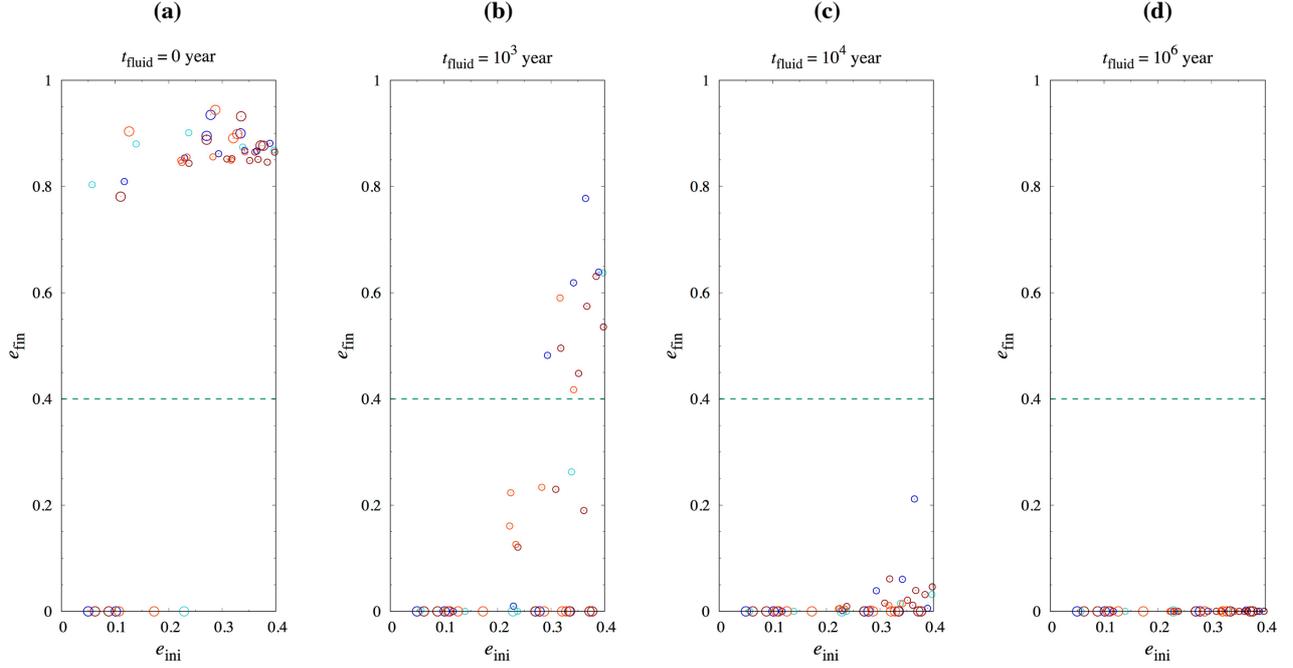

**Figure 4.** Final eccentricity after 4.5 Gyr-tidal evolution ($e_{\text{fin}}$) for different times taken for intact moons to become rigid bodies ($t_{\text{fluid}}$). The size of each circle represents the $\gamma_{\text{sp}}$ value, as in Fig. 3. The four cases were as follows: (a) $t_{\text{fluid}} = 0$ years, in which planetary bodies were rigid for the whole time; (b) $t_{\text{fluid}} = 10^3$ years, in which represents that planetary bodies behaved as fluid for the first $10^3$ years after the giant impact, before becoming rigid; (c) $t_{\text{fluid}} = 10^4$ years; and (d) $t_{\text{fluid}} = 10^6$ years. The tidal evolution pathways with $e_{\text{ini}}$ values larger than ~ 0.4 could not be precisely calculated for the semi-analytical method that we used; a higher order method in $e$ should be used (see ref. [21]). Therefore, in this study we only calculated tidal evolution where $e_{\text{ini}}$ was lower than 0.4.

If intact moons were rigid immediately from their formation point, in our simulations they most likely turned into eccentric satellites. Even when satellites were in a fluid-like state for the first $10^3$ years, almost all satellites with $\gamma_{\text{sp}}$ values lower than $10^{-1}$ became eccentric. These simulations seem to be inconsistent with observations (Supplementary Table 1). In contrast, if the duration of the initial fluid-like state was longer than $10^4$ years, our calculations suggested that almost all intact moons can turn into circular satellites.

Figure 5 shows the final spin and orbital periods of satellite systems formed as intact moons. Among the satellites around large TNOs, only the Pluto-Charon system is in a dual-synchronous state, i.e., the spin periods of satellite ($P_{\text{spin,s}}$) and the primary ($P_{\text{spin,p}}$) coincide with the orbital period ($P_{\text{orb}}$). Note that the spin of Hi'iaka, the largest satellite of Haumea, is not synchronized with its orbital period (see Supplementary Section 3). This strongly implies that the current satellites around Haumea might not have formed near Haumea: they may instead have formed far from its primary. One possibility is that Hi'iaka and Namaka were formed via catastrophic disruption and re-accretion of a past moon which was located near the current orbits of the two satellites[11,22], implying that the Haumean system experienced at least two giant impact events.

We found that the current spin and orbital periods of largest TNOs could be well explained when assuming planetary bodies would behave as fluid-like bodies for $10^4$–$10^6$ years after a giant impact. If the duration of fluid-like behavior was $4.5 \times 10^9$ years, however, most of the satellite systems reached a dual-synchronous state, and there were no satellites with spin and orbital periods are similar to the Eris-Dysnomia or Quaoar-Weywot systems.

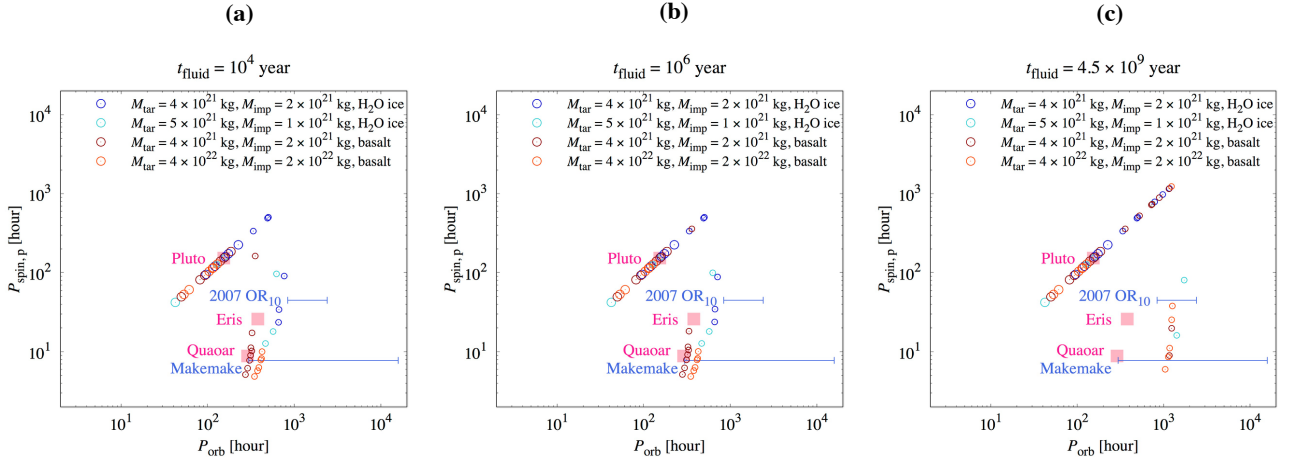

**Figure 5.** Final $P_{\text{spin,p}}$ and $P_{\text{orb}}$ values. The observational data of $P_{\text{spin,p}}$ and $P_{\text{orb}}$ are represented by pink square markers and blue bars, respectively. The size of each circle represents $\gamma_{\text{sp}}$, as in Fig. 3. Case (a) shows the outcome of tidal evolution when planetary bodies were fluid for the first $10^4$ years and became rigid from $10^4$ years to $4.5 \times 10^9$ years after a giant impact. Case (b) shows the outcome of tidal evolution when planetary bodies were fluid for the first $10^6$ years and became rigid from $10^6$ years to $4.5 \times 10^9$ years after a giant impact. Case (c) shows the outcome of tidal evolution when planetary bodies were fluid for the whole $4.5 \times 10^9$ year duration.

We can give some constraints on the thermal evolution of large TNOs from these results. For the case of giant impacts with 1000 km-sized bodies, the effects of impact heating on the internal thermal state of planetary bodies are limited (e.g., ref. [3]). The heat generation from tidal heating can be calculated using the orbital energy of the satellite and the tidal evolution timescale (Supplementary Section 2). We found that, when satellites were initially in solid-state, they could not heat up enough to melt, even if their semimajor axis was as small as their Roche radius.

These results suggest that if all satellites around TNOs have a circular orbit, they were fully or partially molten during the giant impact era. This is consistent with the results of a previous study[23], which suggests that Charon and other small satellites around Pluto were formed via giant impacts with partially differentiated progenitors.

Experimental studies of dust aggregation (e.g., ref. [24]) suggest that micron-sized dust grains aggregate into centimeter-size pebbles in the gaseous solar nebula. In addition, observation of Comet 103P/Hartley[25] shows that its dusty coma is made up of centimeter-sized pebbles. Therefore, kilometer-sized planetesimals with abundant centimeter-sized pebbles may be present at the onset of dwarf planet formation. A recent study[16] revealed that large TNOs can be formed in situ within a period of a few million years, from a solid belt of kilometer-sized and centimeter-sized bodies. Or, when large TNOs form via accretion of pebbles onto planetesimals in a gaseous solar nebula[26], the accretion timescale must be shorter than the lifetime of the solar nebula (~ 4 Myr: ref. [27]). In these cases, large TNOs that form within a few million years can enter either a partially or fully molten state[13]. In addition, when the collision timescale of gravitationally focused large TNOs is also of the order of 10 Myr, then this implies that satellite formation around the TNOs occurred before the outward migration of the outer planets including Neptune[4]. This migration is related to the Late Heavy Bombardment (~ 700 Myr after the formation of the solar system[28]), which is thought to have been a favorable epoch for the formation of rings and inner regular satellites around Saturn, Uranus, and Neptune[29].

Our study shows that satellites around large TNOs may be intact moons formed via giant impacts, similar to the impact that formed Charon. Our simulated secondary-to-primary mass ratios varied over a wide range, and overlapped with observed mass ratios for large TNOs. We also revealed that the current spin/orbital periods distribution and small eccentricity of the satellite systems could be explained only when we considered the tidal evolution of bodies to be fluid-like for at least for the first $10^4$ years. This thermal history is consistent with the current understanding of the formation of dwarf planets in the outer solar system.

**Acknowledgements**

We are grateful to Taishi Nakamoto for his fruitful comments. This work is supported by the Astrobiology Center Program of National Institutes of Natural Sciences (JY290107). S.A. is supported by the Grant-in-Aid for JSPS Research Fellow (JP17J06861). H.G. is supported by MEXT KAKENHI grant (JP17H06457). R.H. acknowledges the financial support of JSPS Grant-in-Aid for JSPS Fellows (JP17J01269) and Early-Career Scientists (JP18K13600).

**Author contributions**

S.A. and H.G. developed the idea for the study and H.G developed the hydrodynamic code for the giant impact simulations. S.A. performed the hydrodynamic simulations and the semi-analytical tidal evolution calculations, with support from R.H. All of the authors contributed to the interpretation of the results.

**Methods**

*1. Numerical code and initial settings for giant impacts*

We used the SPH method to perform giant impact simulations, (e.g., ref. [30]). The SPH method can easily trace large deformations and shock waves, and has been used in many previous giant impact simulations. Our numerical code is the same as that used in Genda et al. [31] and it can calculate a purely hydrodynamic flow with self-gravity, but without material strength.

We considered three types of pre-impact bodies; undifferentiated basaltic bodies, undifferentiated icy bodies, and differentiated bodies with a basalt core and an icy mantle. We applied the Tillotson equation of state[32], which is widely used for giant impact simulations, to calculate the pressure from the internal energy and the density. We note that some previous studies assumed other materials for differentiated/undifferentiated planetary bodies. For example, Canup[3] used a hydrated silicate, serpentine, for the material of undifferentiated Pluto-Charon progenitors, whereas Sekine et al.[13] assumed well-mixed mixtures of $H_2O$ ice and basalt as the materials of Pluto-Charon progenitors. A detailed study on the dependence of the outcome of collision on the material chosen should be performed in the future.

For our simulations, all SPH particles in the planetary bodies were set to have an equal mass ($m_0$) and the number of total particles, $N_{total}$ = 24000. Two impactor-to-target mass ratios ($\gamma_{imp}$) were used in this study, 1/2 and 1/5; the numbers of particles in the target and impactor for these ratios were 16000 and 8000 for $\gamma_{imp}$ = 1/2, and 20000 and 4000 for $\gamma_{imp}$ = 1/5. The internal energy of the SPH particles was set to $5.0 \times 10^5$ J kg$^{-1}$ and neither the impactor nor the target had pre-impact spin. The method for placement of SPH particles is detailed in Genda et al. [14].

We prepared 434 sets of initial conditions for the giant impact simulations. The parameters used in this study were the total mass of the target and the impactor ($M_{total}$), the impactor-to-target mass ratio ($\gamma_{imp}$,), composition and differentiated state, the impact angle ($\theta_{imp}$), and the impact velocity ($v_{imp}$) We considered two different values for the total mass, $6 \times 10^{21}$ kg and $6 \times 10^{22}$ kg.

For each case, we varied $\theta_{imp}$ in 15° steps from 30° to 75°, and the $v_{imp}$ in $0.1v_{esc}$ steps from $0.9v_{esc}$ to $1.7v_{esc}$, above the hit-and-run criteria, where $v_{esc}$ is the two-body escape velocity[33]. In order to precisely determine the intact-moon-forming parameter space, we varied $v_{imp}$ with much smaller steps ($0.02v_{esc}$) below the hit-and-run criteria.

The impact parameters $v_{imp}$ and $\theta_{imp}$ are defined when the two planetary bodies are in contact with each other. To prepare the initial condition, we back-calculated the positions of the two planetary bodies until their distance from each other was $3(R_{tar} + R_{imp})$ for the cases of $v_{imp} < v_{esc}$ and at $10(R_{tar} + R_{imp})$ for the cases of $v_{imp} \geq v_{esc}$, where $R_{tar}$ and $R_{imp}$ are the radius of the target and impactor, respectively, assuming planetary bodies are mass points (see Fig. 1 of Genda et al. [14]). We then performed the SPH simulations over a period of $10^6$ s.

*2. Analysis of the masses of planetary bodies*

To obtain the masses of planetary bodies from the collision outcome data, we used a friends-of-friends algorithm[14] to identify clumps of SPH particles. If the distance between two particles was less than a critical value ($l_{FOF}$) we defined these particles as belonging to the same clump. $l_{FOF}$ is given by $l_{FOF} = \sqrt{3}(m_0/\rho_0)^{1/3}$, where $m_0 = M_{total}/N_{total}$ is the mass of each SPH particle, and $\rho_{0,basalt}$ = 2700 kg m$^{-3}$ and $\rho_{0,ice}$ = 917 kg m$^{-3}$ are the typical densities of the basaltic and icy solid bodies[32]. The value of $l_{FOF}$ is slightly larger than the typical distance between two nearest SPH particles under initial conditions. The distances for all pairs of particles were calculated and we identified clumps.

We regarded clumps that had more than 24 particles as planetary bodies. We termed the largest planetary body the "primary", and the second largest body the "secondary". There were also SPH particles that did not belong to any planetary bodies after collision. We termed these particles isolated particles, and if isolated particles were gravitationally bound to the primary, we identified these particles as disk particles.

In this paper, we mainly focus on the formation of intact moons, but we also discuss satellite formation via impact-generated disks. The dependence of the mass of the disk-origin moon ($M_{DM}$) on the disk mass ($M_{disk}$) is given by Hyodo et al. [18] and references therein.

*3. Orbital Evolution of Formed Satellites*

Both the planet and satellite raise tides on each other. The tidal lag caused by friction leads to angular momentum exchange, which also leads to spin and orbital evolution. The tidal evolution of Pluto-Charon has been investigated by several previous studies[34,35]. In this study, we used $O(e^6)$ tidal evolution equations[36] combined with the constant phase lag model[37] (see Supplementary Section 1 for details). Note that if we want to calculate the tidal evolution pathways whose eccentricities are larger than ~ 0.4, we must consider the eccentricity $e$ for a higher order[37]. In this study, however, we only calculated tidal evolution where the initial eccentricity was lower than 0.4.

**Methods References**

Supplementary Information for:

# Early formation of moons around large trans-Neptunian objects via giant impacts

Sota Arakawa[1], Ryuki Hyodo[2], and Hidenori Genda[2]

[1]Department of Earth and Planetary Sciences, Tokyo Institute of Technology, Meguro, Tokyo, 152-8551, Japan

[2]Earth-Life Science Institute, Tokyo Institute of Technology, Meguro, Tokyo, 152-8550, Japan

## Supplementary Sections

*1. THE EQUATIONS OF TIDAL EVOLUTION*

Tides are raised on both the primary and secondary. In this study, we used the $O(e^6)$ tidal evolution equations[1] combined with the constant phase lag model[2]. The orbit-averaged variations of the spin rates of the planet, $\Omega_p$, and satellite, $\Omega_s$, the semimajor axis, $a$, and the eccentricity, $e$, are given by:

$$-\frac{1}{n}\frac{d\Omega_p}{dt} = \frac{15n}{4}\frac{M_s^2}{(M_p+M_s)M_p}\left(\frac{R_p}{a}\right)^3 \frac{k_{2,p}}{Q_p}\sum_{i=-3}^{+3} E_{2,i}^2(e)\,\text{sgn}[2\Omega_p - (2-i)n],$$

$$-\frac{1}{n}\frac{d\Omega_s}{dt} = \frac{15n}{4}\frac{M_p^2}{(M_p+M_s)M_s}\left(\frac{R_s}{a}\right)^3 \frac{k_{2,s}}{Q_s}\sum_{i=-3}^{+3} E_{2,i}^2(e)\,\text{sgn}[2\Omega_s - (2-i)n],$$

$$\frac{1}{a}\frac{da}{dt} = \frac{3n}{2}\frac{M_s}{M_p}\left(\frac{R_p}{a}\right)^5 \frac{k_{2,p}}{Q_p}\sum_{i=-3}^{+3}(2-i)\,E_{2,i}^2(e)\,\text{sgn}[2\Omega_p - (2-i)n]$$

$$+ \frac{3n}{2}\frac{M_p}{M_s}\left(\frac{R_s}{a}\right)^5 \frac{k_{2,s}}{Q_s}\sum_{i=-3}^{+3}(2-i)\,E_{2,i}^2(e)\,\text{sgn}[2\Omega_s - (2-i)n],$$

$$-\frac{1}{e}\frac{de}{dt} = \frac{3n}{4e^2}\frac{M_s}{M_p}\left(\frac{R_p}{a}\right)^5 \frac{k_{2,p}}{Q_p}\sum_{i=-3}^{+3}\left[2\sqrt{1-e^2}-(2-1)(1-e^2)\right] E_{2,i}^2(e)\,\text{sgn}[2\Omega_p - (2-i)n]$$

$$+ \frac{3n}{4e^2}\frac{M_p}{M_s}\left(\frac{R_s}{a}\right)^5 \frac{k_{2,s}}{Q_s}\sum_{i=-3}^{+3}\left[2\sqrt{1-e^2}-(2-1)(1-e^2)\right] E_{2,i}^2(e)\,\text{sgn}[2\Omega_s - (2-i)n],$$

where $n$ is the mean motion, $M_p$, $M_s$, $R_p$, and $R_s$ are the masses and radii of the primary and the secondary, respectively, and $Q_p$, $Q_s$, $k_{2,p}$, and $k_{2,s}$ are the dissipation functions and Love numbers of the primary and the secondary, respectively. These equations preserve the total angular momentum[1]. In this study, we assumed $Q_p = Q_s = 100$ (ref. [3]). The Love number $k_{2,i}$ is given by,

$$k_{2,i} = \frac{3}{2}\left(1 + \frac{38\pi}{3}\frac{\mu_i R_i^4}{GM_i^2}\right)^{-1},$$

where $\mu_i$ is the rigidity of body $i$ (p for primary and s for secondary), and $G$ is the gravitational constant. We adopted a $\mu_i$ value of $6.5 \times 10^{10}$ Pa for rigid basalt[4] and a $\mu_i$ value of $4 \times 10^9$ Pa for rigid ice[5]. We set $\mu_i = 0$ when the planetary bodies were exhibiting fluid-like behavior, for simplicity. This treatment was used in the tidal evolution calculation of the Eris-Dysnomia system[6]. The eccentricity functions $E_{2,i}(e)$ are presented by Efroimsky[7].

Note that to calculate the tidal evolution pathways with eccentricities where $e \gtrsim 0.4$, we must use the tidal evolution equations that consider the higher-order effects of eccentricity (ref. [2]). In this study, however, we only calculated tidal evolution where the initial eccentricity was lower than 0.4.

In this study, we did not consider the effects of a dynamical tide. The dynamical tide can dump eccentricity efficiently when the periapsis distance is small, however, the current dynamical tide model is not suitable for rigid planetary bodies, and this is beyond the scope of this study.

When the primary rotates quickly and the secondary is in synchronous state, i.e., $\Omega_p \gg (3/2)n$ and $\Omega_s = n$, the tidal torque caused by the primary increases the orbital eccentricity, while the tidal torque caused by the secondary decreases eccentricity. In this case, the orbit-averaged variation of eccentricity can be simplified as follows:

$$\frac{1}{e}\frac{de}{dt} \simeq \frac{33n}{4}\frac{M_s}{M_p}\left(\frac{R_p}{a}\right)^5 \frac{k_{2,p}}{Q_p}\left(1 - \frac{25}{22}A\right),$$

where:

$$A = \frac{k_{2,s}}{k_{2,p}}\left(\frac{Q_s}{Q_p}\right)^{-1}\left(\frac{M_s}{M_p}\right)^{-2}\left(\frac{R_s}{R_p}\right)^5.$$

The dimensionless parameter $A$ is the relative rate of tidal dissipation in the planet and satellite[8]. Whether eccentricity increases through tidal evolution can then be determined using the relative rate of tidal dissipation in the planet and satellite; if $A$ is larger than unity, dissipation mainly occurs in the satellite, and eccentricity will decrease[3].

If we assume that the density and the dissipation function are the same for both the primary and the secondary, the parameter $A$ depends on ratios of the planetary radii ($R_s/R_p$) and the Love numbers ($k_{2,s}/k_{2,p}$).

For rigid bodies, the Love number of 1000 km-sized solid-like bodies is smaller than unity, and is proportional to the square of the planetary radius. In contrast, for fluid-like bodies (i.e., partially or fully molten bodies), the Love number is approximately 3/2, and does not depend on the planetary radius.

When the primary and the secondary are both in a rigid state, the relative rate $A$ is:

$$A \simeq \frac{R_s}{R_p} < 1,$$

whereas when both the primary and the secondary are in fluid-like state:

$$A \simeq \frac{R_p}{R_s} > 1.$$

We found that eccentricity will decrease when both the primary and the secondary are in a fluid-like state.

## 2. TIDAL HEATING ON INTACT MOONS

When the primary rotates quickly and the secondary is in synchronous state, the orbit-averaged variation of the semimajor axis can be simplified as follows:

$$\frac{1}{a}\frac{da}{dt} = \frac{3n}{2}\frac{M_s}{M_p}\left(\frac{R_p}{a}\right)^5 \frac{k_{2,p}}{Q_p}\left(2 + 27e^2 + \mathcal{O}(e^4)\right) - \frac{3n}{2}\frac{M_p}{M_s}\left(\frac{R_s}{a}\right)^5 \frac{k_{2,s}}{Q_s}\left(\frac{73}{2}e^2 + \mathcal{O}(e^4)\right).$$

Here we discuss whether satellites in solid-state can turn into fluid-like bodies through tidal heating. When both the primary and the secondary are in a rigid state, the semimajor axis, $a$, increases with the timescale, $\tau_a$, which is given by:

$$\tau_a = \frac{1}{3n}\frac{M_p}{M_s}\left(\frac{a}{R_p}\right)^5 \frac{Q_p}{k_{2,p}} \sim 1 \times 10^5 \left(\frac{a/R_p}{4}\right)^{13/2} \left(\frac{R_s/R_p}{0.2}\right)^{-2} \left(\frac{R_p}{500 \text{ km}}\right)^{-2} \text{ years}.$$

The cooling timescale of a satellite due to heat conduction, $\tau_{\text{cool}}$, is given by,

$$\tau_{\text{cool}} = \frac{\rho c R_s^2}{\lambda} \sim 3 \times 10^8 \left(\frac{R_s}{100 \text{ km}}\right)^2 \text{ years},$$

where $c$ is the specific heat[9] ($c = 1.13 \times 10^3$ J kg$^{-1}$ K$^{-1}$) and $\lambda$ is the thermal conductivity[10] ($\lambda = 2.1$ W K$^{-1}$ m$^{-1}$). In comparison with the cooling timescale and the tidal dissipation timescale, we found that most of the thermal energy generated by tidal dissipation is retained in both the primary and the secondary, which can increase the internal temperature of satellites.

The increase of temperature is given by $\Delta T_s \sim \Delta E_s/(M_s c)$, where $\Delta E_s$ is the thermal energy generated by tidal dissipation in the satellite. The thermal energy, $\Delta E_s$, can be estimated from $\Delta E_s = \dot{E}_s \tau_a$, where $\dot{E}_s$ is the heat generation in the satellite per unit time. The heat generation rate, $\dot{E}_s$, is given by:

$$\dot{E}_s = \frac{1}{2}\frac{GM_pM_s}{a}\tau_e^{-1},$$

where the energy dissipation timescale of the satellite, $\tau_e$, is given by:

$$\tau_e = \frac{4}{219 e^2 n}\frac{M_s}{M_p}\left(\frac{a}{R_s}\right)^5 \frac{Q_s}{k_{2,s}}.$$

Therefore, the increase of the internal temperature of the satellite, $\Delta T_s$, is given by:

$$\Delta T_s \simeq \frac{73 e^2}{8}\frac{R_s}{R_p}\frac{GM_p}{ac} \sim 5 \left(\frac{a/R_p}{4}\right)^{-1} \left(\frac{e}{0.4}\right)^2 \left(\frac{R_s/R_p}{0.2}\right) \left(\frac{R_p}{500 \text{ km}}\right)^2 \text{ K}.$$

Therefore, the effect of tidal dissipation is limited, and it is unlikely that satellites can move into a fluid-like state through tidal heating alone.

*3. CATASTROPHIC DISRUPTION OF PRIMORDIAL MOONS AND THE FORMATION OF HAUMEAN MOONS*

Haumea has a short spin period of 3.92 hours[11], and it has two satellites named Hi'iaka and Namaka. The masses of Haumea, Hi'iaka, and Namaka are $4.0 \times 10^{21}$ kg, $1.8 \times 10^{19}$ kg, and $1.8 \times 10^{18}$ kg, respectively[12], and the secondary-to-primary mass ratio ($\gamma_{sp}$) of the Haumean system is $\sim 4.5 \times 10^{-3}$. If these two satellites were formed close to Haumea, their spins would be expected to have been tidally de-spun, and they should therefore be rotating synchronously with their own orbital periods. The spin period of Hi'iaka is about 120 times faster than its orbital period[13], however, suggesting that the two current satellites were formed via the catastrophic disruption and re-accretion of a past moon that was located near to the current orbits of the two satellites[14,15]. This mechanism has also been debated in the context of the late formation of Saturn's midsized moons and its rings[16,17].

# Supplementary Figures

**(a) Formation of an intact moon without a rocky core**

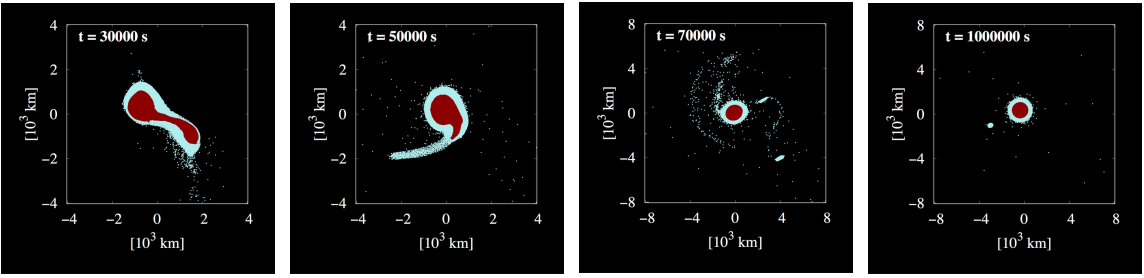

**(b) Disk formation without an intact moon**

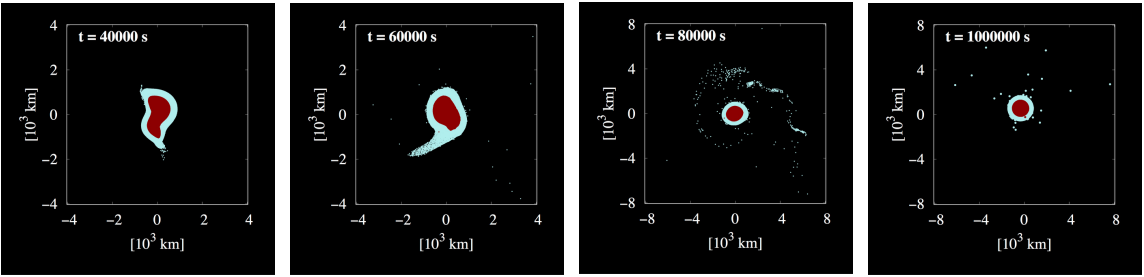

**(c) Hit-and-run collision**

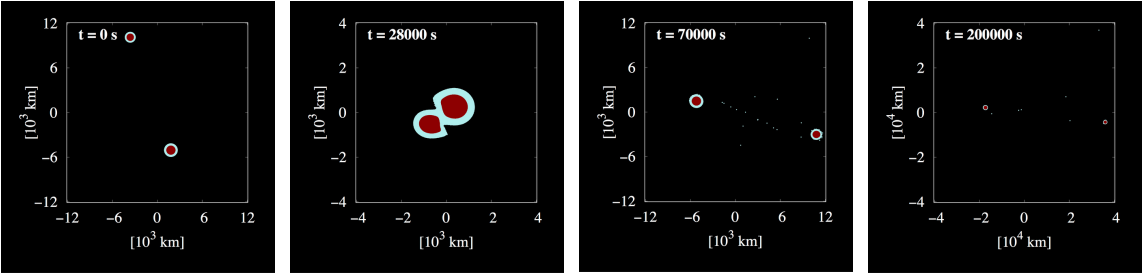

**Supplementary Figure 1.** Snapshots of giant impacts between two differentiated bodies. Both the target and the impactor were differentiated bodies with $f_{ice}$ = 0.5 (50 wt.% ice and 50 wt.% basalt). The target mass, $M_{tar}$, and the impactor mass, $M_{imp}$, are $4 \times 10^{21}$ kg and $2 \times 10^{21}$ kg, respectively. (a) the case for $v_{imp} = 1.25 v_{esc}$ and $\theta_{imp}$ = 30°, resulting in formation of an intact moon without a rocky core ($\gamma_{sp} = 1.0 \times 10^{-2}$). (b) the case for $v_{imp} = 1.05 v_{esc}$ and $\theta_{imp}$ = 30°, resulting in the formation of a disk without an intact moon ($M_{disk} / M_p = 3.6 \times 10^{-3}$ and the estimated mass of the disk-origin moon is $M_{DM} / M_p = 9.0 \times 10^{-5}$). (c) the case for $v_{imp} = 1.20 v_{esc}$ and $\theta_{imp}$ = 60°, resulting in a hit-and-run collision.

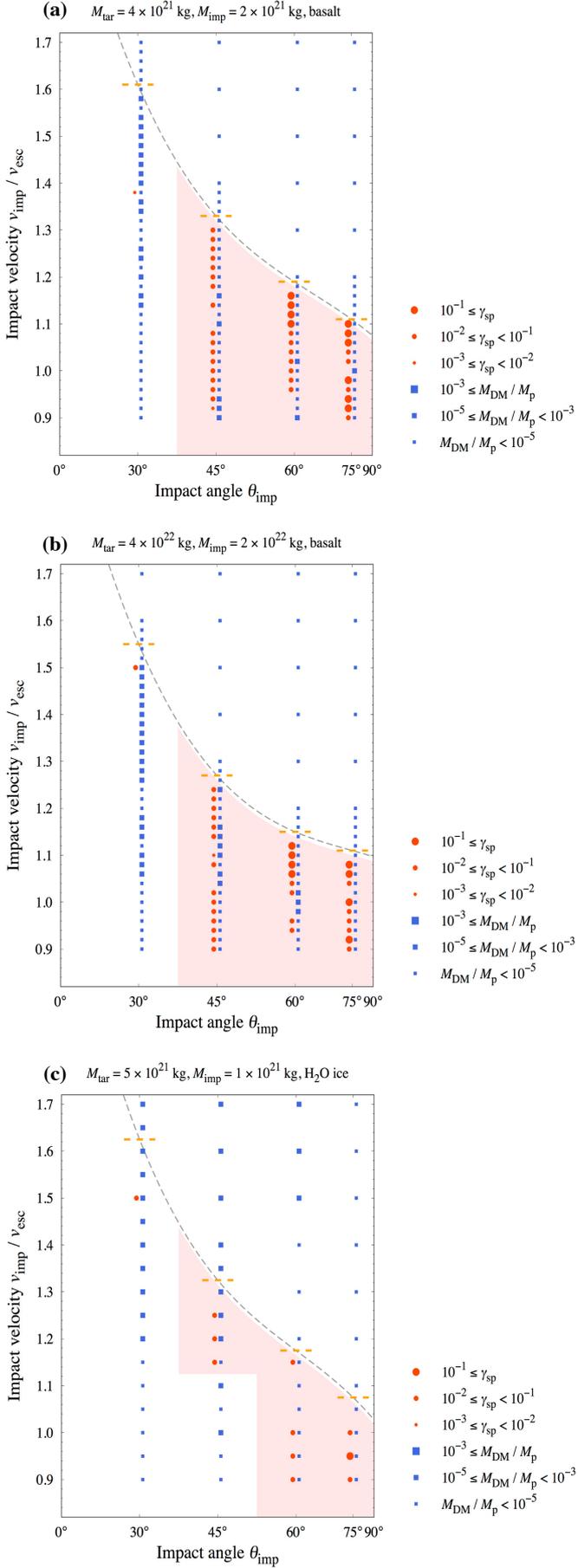

**Supplementary Figure 2.** Summary of the range of outcomes for giant impacts (as Fig. 2 in the main text). (a) outcomes for giant impacts between rocky undifferentiated planetary bodies with masses of $4 \times 10^{21}$ kg and $2 \times 10^{21}$ kg. (b) outcomes for giant impacts between rocky undifferentiated planetary bodies with masses of $4 \times 10^{22}$ kg and $2 \times 10^{22}$ kg. (c) outcomes for giant impacts between icy undifferentiated planetary bodies with masses of $5 \times 10^{21}$ kg and $1 \times 10^{21}$ kg.

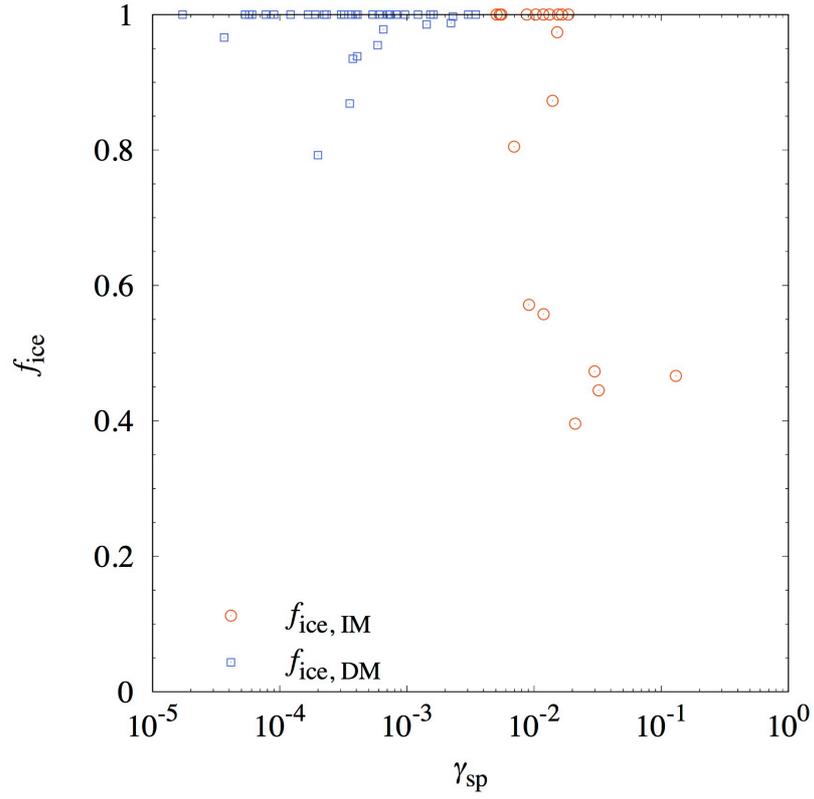

**Supplementary Figure 3.** The ice mass fraction of intact moons ($f_{ice,IM}$; red circles) vs. the secondary-to-primary mass ratio $\gamma_{sp}$. Note that, for cases where multiple intact moons were formed, only the $f_{ice,IM}$ of the largest intact moon is shown. We also plot the ice mass fraction of disk-origin moons ($f_{ice,DM}$; blue squares) vs. the estimated $\gamma_{sp}$ of the largest disk-origin moon. IM = intact moon, DM = disk-origin moon.

# Supplementary Table

| primary ($R_p$) | secondary ($R_s$) | $\gamma_{sp}$ | $P_{spin,p}$ [h] | $P_{spin,s}$ [h] | $P_{orb}$ [h] | $e$ |
|---|---|---|---|---|---|---|
| **Pluto (1187 km)** | **Charon (606 km)** | $1.2 \times 10^{-1}$ | **153.2** | **153.2** | **153.2** | $\sim 5 \times 10^{-5}$ |
| (ref. [18]) | (ref. [18]) | (ref. [19]) | (ref. [19]) | (ref. [19]) | (ref. [19]) | (ref. [19]) |
| **Eris (1163 km)** | **Dysnomia (350 km)** | $3 \times 10^{-2}$ | **25.9** | (unknown) | **379** | $< 4 \times 10^{-3}$ |
| (ref. [20]) | (ref. [21]) | | (ref. [22]) | | (ref. [21]) | (ref. [21]) |
| **Haumea (816 km)** | **Hi'iaka (150 km)** | $4.5 \times 10^{-3}$ | **3.91** | **9.8** | **1187** | $\sim 5 \times 10^{-2}$ |
| (ref. [23]) | (ref. [12]) | (ref. [12]) | (ref. [11]) | (ref. [13]) | (ref. [12]) | (ref. [12]) |
| **Makemake (715 km)** | **(no name) (90 km)** | $2 \times 10^{-3}$ | **7.77** | (unknown) | **300−16000** | (unknown) |
| (ref. [24]) | (ref. [25]) | | (ref. [26]) | | (ref. [25]) | |
| **2007 OR10 (770 km)** | **(no name) (180 km)** | $1 \times 10^{-2}$ | **44.8** | (unknown) | **840−2400** | (unknown) |
| (ref. [27]) | (ref. [28]) | | (ref. [27]) | | (ref. [28]) | |
| **Quaoar (535 km)** | **Weywot (40 km)** | $4 \times 10^{-4}$ | **8.84** | (unknown) | **289.0** | $\sim 0.15$ |
| (ref. [29]) | (ref. [29]) | | (ref. [30]) | | (ref. [31]) | (ref. [31]) |

**Supplementary Table 1.** Radii of primary and secondary ($R_p$ and $R_s$), secondary-to-primary mass ratio ($\gamma_{sp}$), spin periods of primary and secondary ($P_{spin,p}$ and $P_{spin,s}$), the orbital period ($P_{orb}$), and the eccentricity of the satellite systems around large trans-Neptunian objects. Note that, except for Pluto-Charon and Haumea-Hi'iaka, we estimated $\gamma_{sp}$ by assuming that both the primary and the secondary have equal densities.